\documentclass[12pt]{article}
\usepackage[a4paper,margin=1in]{geometry}
\usepackage[titletoc,title]{appendix}
\usepackage{changepage}
\usepackage[utf8x]{inputenc}
\usepackage{textcomp,marvosym}
\usepackage{fixltx2e}
\usepackage{amsmath,amssymb}
\usepackage{cite}
\usepackage{nameref,hyperref}
\usepackage{authblk}
\usepackage{algorithm}
\usepackage{algpseudocode}
\usepackage{dsfont}
\usepackage{xcolor}
\usepackage{enumitem}
\usepackage{caption,setspace}
\usepackage{subcaption}
\usepackage{pdflscape}
\usepackage{graphicx}
\usepackage{placeins}
\usepackage{mathtools}

\renewcommand{\eqref}[1]{Equation~(\ref{#1})}

\def\cbl{\color{black}}
\def\cb{\color{black}}

\title{Mean exit time for diffusion on irregular domains}

\author[1]{Matthew~J. Simpson\footnote{To whom correspondence should be addressed. E-mail: matthew.simpson@qut.edu.au}}
\author[1]{Daniel~J. Vandenheuvel}
\author[1]{Joshua~M. Wilson}
\author[1]{Scott~W. McCue}
\author[1]{Elliot~J. Carr}
\affil[1]{School of Mathematical Sciences, Queensland University of Technology, Brisbane, Queensland 4001, Australia}

\renewcommand{\epsilon}{\varepsilon}

\newcommand{\matlab}{MATLAB}
\begin{document}

\maketitle
\begin{abstract}
Many problems in physics, biology, and economics depend upon the duration of time required for a diffusing particle to cross a boundary.  As such, calculations of the distribution of first passage time, and in particular the mean first passage time, is an active area of research relevant to many disciplines.  Exact results for the mean first passage time for diffusion on simple geometries, such as lines, discs and spheres, are well--known.  In contrast, computational methods are often used to study the first passage time for diffusion on more realistic geometries where closed--form solutions of the governing elliptic boundary value problem are not available.  Here, we develop a perturbation solution to calculate the mean first passage time on irregular domains formed by perturbing the boundary of a disc or an ellipse. Classical perturbation expansion solutions are then constructed using the exact solutions available on a disc and an ellipse.  We apply the perturbation solutions to compute the mean first exit time on two naturally--occurring irregular domains: a map of Tasmania, an island state of Australia, and a map of Taiwan.  Comparing the perturbation solutions with numerical solutions of the elliptic boundary value problem on these irregular domains confirms that we obtain a very accurate solution with a few terms in the series only.   \matlab\ software to implement all calculations is available at \href{https://github.com/ProfMJSimpson/Exit_time}{https://github.com/ProfMJSimpson/Exit\_time}.
\end{abstract}


\newpage
\section{Introduction} \label{intro}
Mathematical models describing diffusive transport phenomena are fundamental tools with broad applications in many areas of physics~\cite{Redner2001,Krapivsky2010,Hughes1995}, engineering~\cite{Bear1972,Bird2002} and biology~\cite{Murray2002,Kot2003,Edelstein2005}.  Traditional analysis of mathematical models of diffusion often focus on idealised problems with relatively simple geometries, whereas practical applications routinely involve more complicated geometries that are normally dealt with using computational approaches.  While computational approaches are essential in many circumstances~\cite{Lotstedt2015,Meinecke2016}, exact analytical insight is preferable, where possible, because it can lead to mathematical expressions that explicitly highlight key relationships that are not always revealed computationally.

A fundamental property of diffusion is the concept of particle lifetime, which is a particular application of the more general concept of the first passage time~\cite{Redner2001,Krapivsky2010,Hughes1995}.  Estimates of particle lifetime provide insight into the timescale required for a diffusing particle to reach a certain target, such as an absorbing boundary.  Many results are known about the particle lifetime for diffusion in simple geometries, such as lines and discs~\cite{Redner2001,Krapivsky2010}.  Generalising these results to deal with other geometrical features, such as wedges~\cite{Di2008,Chupeau2015}, symmetric domains~\cite{Vaccario2015,Rupprecht2015,Carr2020}, growing domains~\cite{Simpson2015a,Simpson2015b}, slender domains~\cite{Kurella2014,Lindsay2015,Grebenkov2019}, small targets~\cite{Lindsay2017,Grebenkov2020} or arbitrary initial conditions~\cite{Nyberg2016} is an active area of research.

In this work, we develop new expressions for particle lifetime for diffusion in irregular two--dimensional (2D) geometries.  Our approach is to  use classical results for diffusion on a disc and an ellipse, and then construct perturbation solutions for more general geometries~\cite{McCue2011,McCollum2014}.  The new perturbation solutions are evaluated using symbolic computation, and the results compare very well with numerical solutions of the governing boundary value problem, and with averaged data from repeated stochastic simulations.  Once the new perturbation solutions are derived and validated, we consider two case studies to show how these tools can be used to take a particular  2D region and represent it as a perturbed ellipse or disc.  This allows us to use the new perturbation solutions to analyse the mean particle lifetime in irregular 2D geometries.  \matlab\ code provided on \href{https://github.com/ProfMJSimpson/Exit_time}{https://github.com/ProfMJSimpson/Exit\_time} is used to compute: (i) the symbolic evaluation of the perturbation solution; (ii) the numerical finite volume solution of the boundary value problem; and, (iii) the stochastic random walk algorithm.

\section{Results and Discussion} \label{Results}

Standard arguments that relate unbiased random walk models on domains with absorbing boundary conditions can be used to show that the mean exit time is given by the solution of a linear ellipse partial differential equation~\cite{Redner2001,Krapivsky2010,Hughes1995}. In this work we seek solutions of that equation,
\begin{align}
D \nabla^2 T(\mathbf{x}) &= -1, \quad \mathbf{x} \in \Omega,  \quad \textrm{with}  \label{eq:GovEq} \\
T &=0, \quad \textrm{on} \quad \partial \Omega, \label{eq:GovEqBoundary}
\end{align}
where $D>0$ is the diffusivity and $T(\mathbf{x})$ is the mean exit time for a particle released at location $\mathbf{x}$.  Here the diffusivity is related to the random walk model, $D = \mathcal{P} \delta^2 / (4 \tau)$, where $\delta>0$ is the step length, $\tau>0$ is the duration between steps and $\mathcal{P} \in [0,1]$ is the probability that the particle undergoing Brownian motion attempts to undergo a step of length $\delta$ in the time duration $\tau$~\cite{Redner2001,Krapivsky2010,Hughes1995}.  \cbl The continuum description is valid in the constrained limit  $\delta \to 0$, $\tau \to 0$ and $\delta^2/\tau$ held constant~\cite{Redner2001,Krapivsky2010,Hughes1995}.  The stochastic random walk model is Pearson's walk in $\mathbb{R}^2$, and full details of how simulations are performed are given in Appendix A. \cb

Key results are presented and discussed in the following format.  In Sections \ref{disc} and \ref{ellipse} we review known exact solutions to Equation (\ref{eq:GovEq}) on a disc and an ellipse, respectively.  In Sections \ref{perturbed_disc} and \ref{perturbed_ellipse} we develop new approximate solutions on irregular domains, and in Section \ref{sec:CaseStudies} we apply the new approximate solutions to study the exit time for diffusion in two naturally--occurring geometries.

\subsection{Mean exit time on a disc} \label{disc}
\begin{figure}[H]
	\centering
	\includegraphics[width=1.0\textwidth]{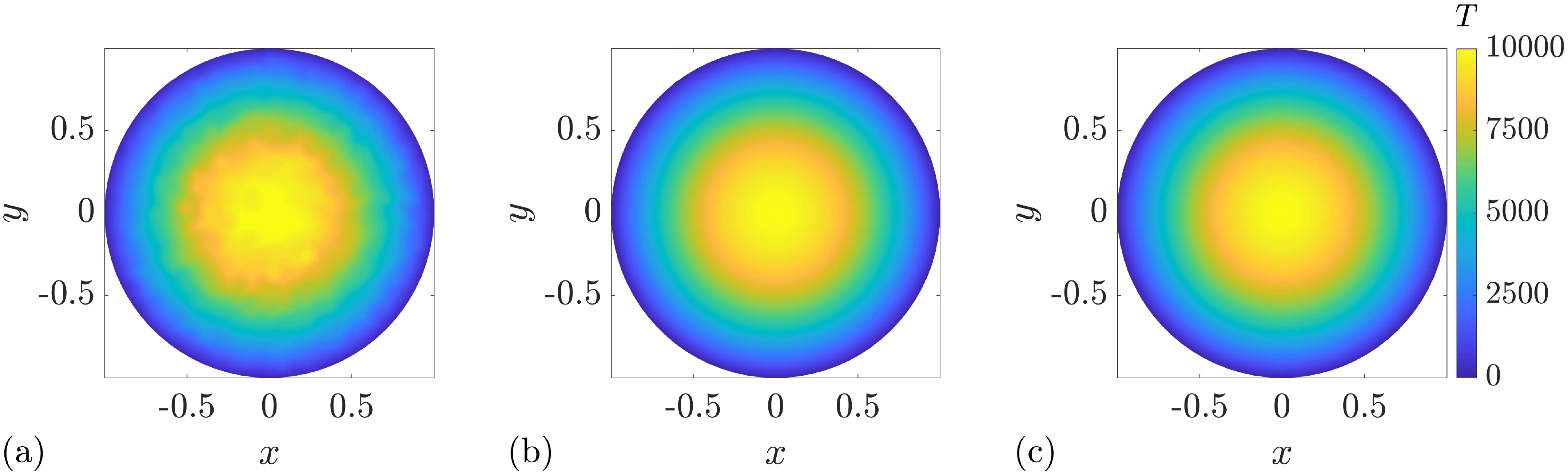}
	\caption{\textbf{Mean exit time on the unit disc.} (a) Averaged data from repeated stochastic simulations. (b) Exact solution of Equations (\ref{eq:GovEq})--(\ref{eq:GovEqBoundary}). (c) Numerical solution of Equations (\ref{eq:GovEq})--(\ref{eq:GovEqBoundary}). Parameters are $\tau=1$, $\mathcal P=1$, $\delta = 1 \times 10^{-2}$ and $D=2.5 \times 10^{-5}$. The triangular mesh used to construct the solution in (a) and (c) has $632$ nodes and $1183$ triangular elements. \cbl For (a), 1000 random walks starting from each node were generated. \cb}
	\label{fig:F1}
\end{figure}

For the special case where $\Omega$ is a disc of radius $R>0$ centered at the origin it is convenient to work in polar coordinates, where the mean first exit time depends upon the radial coordinate only,
\begin{equation}
\dfrac{D}{r}\dfrac{\text d}{\text dr}\left[ r \dfrac{\text d T(r)}{\text d r}\right] = -1, \quad 0 < r < R. \label{eq:GovEqdisc}
\end{equation}
The solution of \eqref{eq:GovEqdisc} with appropriate boundary conditions, $T(R)=0$ and $\textrm{d}T(0)/\textrm{d}r=0$, is given by
\begin{equation}
T(r) = \dfrac{R^2-r^2}{4D}. \label{eq:Solutiondisc}
\end{equation}
Results in Figure \ref{fig:F1}(a)--(c) provide a visual comparison of exit time data from repeated stochastic simulations, the exact solution and the numerical solution of Equations (\ref{eq:GovEq})--(\ref{eq:GovEqBoundary}) for a unit disc, $R=1$.  A complete description of the continuous space, discrete time stochastic algorithm and the finite volume numerical method used with an unstructured triangular mesh to solve \eqref{eq:GovEq} is given in Appendix A.

\subsection{Mean exit time on an ellipse} \label{ellipse}
\begin{figure}[H]
	\centering
	\includegraphics[width=1.0\textwidth]{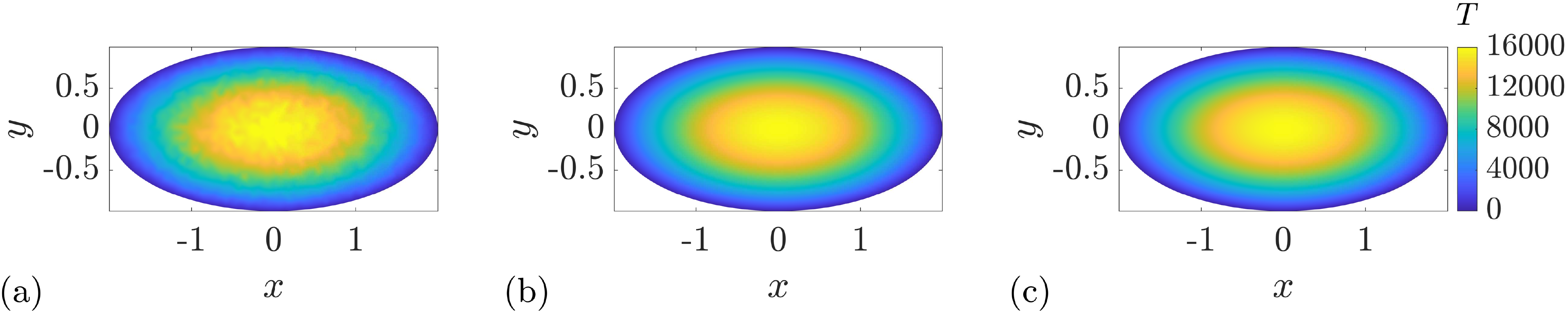}
	\caption{\textbf{Mean exit time on an ellipse with $a=2$ and $b=1$.} (a) Averaged data from repeated stochastic simulations. (b) Exact solution of \eqref{eq:GovEq}. (c) Numerical solution of Equations (\ref{eq:GovEq})--(\ref{eq:GovEqBoundary}). Parameters are $\tau=1$, $\mathcal P=1$, $\delta = 1 \times 10^{-2}$ and $D=2.5 \times 10^{-5}$. The triangular mesh used to construct the solution in (a) and (c) has $1240$ nodes and $2356$ triangular elements. \cbl For (a), 1000 random walks starting from each node were generated. \cb}
	\label{fig:F2}
\end{figure}

For the special case where $\partial \Omega$ is an ellipse centered at the origin it is convenient to work in Cartesian coordinates.  The  ellipse, given by
\begin{equation}
\dfrac{x^2}{a^2} + \dfrac{y^2}{b^2} = 1, \label{eq:EllipseBoundary}
\end{equation}
has width $2a > 0$ and height $2b > 0$.  The solution of \eqref{eq:GovEq} on this domain is \cbl given by~\cite{McCue2011} \cb
\begin{equation}
T(x,y)= \dfrac{a^2b^2}{2D\left(a^2+b^2\right)}\left[1- \dfrac{x^2}{a^2} - \dfrac{y^2}{b^2}\right]. \label{eq:EllipseSolution}
\end{equation}
Results in Figure \ref{fig:F2}(a)--(c) provide a visual comparison of exit time data from repeated stochastic simulations, the exact solution and the numerical solution of \eqref{eq:GovEq} for an ellipse with $a=2$ and $b=1$.

\subsection{Mean exit time on a perturbed disc} \label{perturbed_disc}
We begin working on irregular domains by calculating expressions for the exit time on a perturbed disc.  \cbl Using plane polar coordinates $(r, \theta)$, we consider a region $\Omega$  with boundary $r = \mathcal{R}(\theta)$, subject to the condition that $\mathcal{R}(\theta)$ is a single-valued function of $\theta$ to ensure that any ray drawn from the origin intersects precisely one point of the boundary $\partial \Omega$.  If our region is conceived as a modest perturbation of a circular disc of radius $R$ we can write
\begin{equation}
\mathcal{R}(\theta) = R\left(1 + \varepsilon g(\theta) \right), \label{eq:DiscPerturbation}
\end{equation}
where $R>0$ is the radius of the unperturbed disc, $\theta \in [0, 2\pi)$ is the polar angle, $\varepsilon \ll 1$ is a small dimensionless parameter and $g(\theta)$ is an $\mathcal O(1)$ smooth periodic function with period $2\pi$.  We assume that the solution can  be written as \cb
\begin{equation}  \label{eq:DiscPowerseries}
T(r, \theta) = T_{0}(r, \theta) + \varepsilon T_{1}(r, \theta) + \varepsilon^2 T_{2}(r, \theta)+ \cdots +\varepsilon^n T_{n}(r, \theta) + \mathcal{O}(\varepsilon^{n+1}).
\end{equation}
When the $\mathcal O(\epsilon^{n+1})$ term is neglected in \eqref{eq:DiscPowerseries} the solution is referred to as the $\mathcal O(\epsilon^n)$ perturbation solution.  To proceed we evaluate $T(r,\theta)$ on $\mathcal{R}(\theta)$ and expand about $\varepsilon=0$ to give,
\begin{align} \nonumber
0 &= T(R + \varepsilon R g(\theta), \theta),\\
0 &= T(R,\theta) + \varepsilon R g(\theta) \left. \dfrac{\partial T}{\partial \theta} \right|_{r=R} + \cdots + \dfrac{\left(\varepsilon R g(\theta)\right)^n}{n!}\left. \dfrac{\partial^n T}{\partial \theta^n} \right|_{r=R} + \mathcal{O}(\varepsilon^{n+1}).\label{eq:DiscBCPowerseries}
\end{align}
Substituting \eqref{eq:DiscPowerseries} into \eqref{eq:DiscBCPowerseries} and equating powers of $\varepsilon$ leads to,
\begin{align} \label{eq:DiscBCs1}
\mathcal{O}(1): \quad &T_{0}(R,\theta)=0,  \\
\mathcal{O}(\varepsilon^\ell): \quad & T_\ell(R, \theta) +  \sum_{k=1}^{\ell} \dfrac{\left(R g(\theta) \right)^k}{k!} \left. \dfrac{\partial^k T_{\ell-k}}{\partial r^k} \right|_{r=R} = 0, \label{eq:DiscBCs2}
\end{align}
for $\ell = 1, \ldots, n$. With this information we substitute \eqref{eq:DiscPowerseries} into \eqref{eq:GovEq} to give a family of boundary value problems for each term in the power series, with the boundary conditions given by Equations (\ref{eq:DiscBCs1})--(\ref{eq:DiscBCs2}).  This family of boundary value problems can be summarised as
\begin{align} \label{eq:DiscBVPs1}
&\mathcal{O}(1): \quad D \nabla^2 T_{0} =-1, \quad T_{0}(R, \theta)=0, \\
&\mathcal{O}(\varepsilon^\ell): \quad \nabla^2 T_{\ell} = 0, \quad T_{\ell}(R, \theta)=-\sum_{k=1}^{\ell} \dfrac{\left(R g(\theta) \right)^k}{k!}  \left. \dfrac{\partial^k T_{\ell-k}}{\partial r^k} \right|_{r=R}, \label{eq:DiscBVPs2}
\end{align}
for $\ell = 1, \ldots, n$. The solution of \eqref{eq:DiscBVPs1} is  \eqref{eq:Solutiondisc}, and each higher order term, $T_{\ell}(r,\theta)$, $\ell = 1,\hdots,n$, is the solution of Laplace's equation on a disc with prescribed boundary data associated with the previous term.  Each higher order term can be obtained using separation of variables, giving
 \begin{equation} \label{eq:DiscBVPSolutions}
T_{\ell}(r,\theta) = \dfrac{a_0}{2} + \sum_{k=1}^{\infty} \left[a_k r^k \cos(k \theta) + b_k r^k \sin(k \theta)  \right],
\end{equation}
\cbl where \cb
\begin{equation}
 a_k = \dfrac{1}{\pi R^k} \int_{0}^{2\pi} T_{\ell}(R,\theta)\cos(k \theta) \, \textrm{d}\theta, \quad b_k = \dfrac{1}{\pi R^k} \int_{0}^{2\pi} T_\ell(R,\theta) \sin(k \theta) \, \textrm{d}\theta,
\end{equation}
for $\ell = 1, \ldots, n$. These solutions are straightforward to evaluate symbolically, and a \matlab\ implementation of the symbolic evaluation of them is available on \href{https://github.com/ProfMJSimpson/Exit_time}{GitHub}.

\begin{figure}[H]
	\centering
	\includegraphics[width=1.0\textwidth]{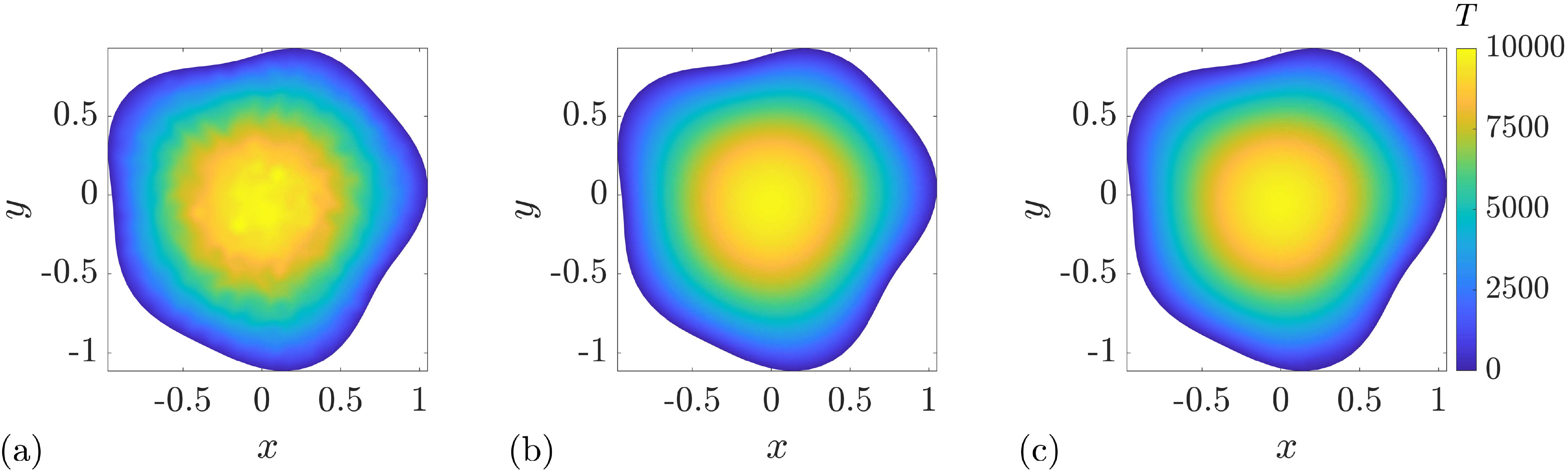}
	\caption{\textbf{Mean exit time on a perturbed disc.} (a) Averaged data from repeated stochastic simulations. (b) $\mathcal{O}(\varepsilon^2)$ perturbation solution. (c) Numerical solution of Equations (\ref{eq:GovEq})--(\ref{eq:GovEqBoundary}).  Parameters are $\tau=1$, $\mathcal P=1$, $\delta = 1 \times 10^{-2}$ and $D=2.5 \times 10^{-5}$. The triangular mesh used to construct the solution in (a) and (c) has $636$ nodes and $1189$ triangular elements. \cbl For (a), 1000 random walks starting from each node were generated. \cb}
	\label{fig:F3}
\end{figure}

Results in Figure \ref{fig:F3}(a)--(c) provide a visual comparison of exit time data from repeated stochastic simulations, the perturbation and numerical solution of \eqref{eq:GovEq} for a perturbed unit disc with $R=1$, $g(\theta) = \sin(3\theta) + \cos(5\theta) - \sin \theta$, and $\varepsilon = 1/20$.  These results show that the truncated perturbation solution is very accurate, despite using only the first three terms in \eqref{eq:DiscPowerseries} and just the first 25 terms to approximate the infinite sum in \eqref{eq:DiscBVPSolutions}.  Perturbation solutions with different choices of truncation, different choices of $\varepsilon$, and  different choices of $g(\theta)$ can be evaluated using the \matlab\ algorithms available on \href{https://github.com/ProfMJSimpson/Exit_time}{GitHub}.  Additional information about how the choice of truncation in \eqref{eq:DiscPowerseries}  affects the accuracy of the perturbation solution is given in the Appendix B.

\cbl It is worth explicitly pointing out some interesting features that arise when we solve \eqref{eq:GovEq} on a perturbed disc.  In sectors where $g(\theta) > 0$, there are points of $\Omega$ that lie beyond the circle $r = R$, whereas in sectors where $g(\theta) < 0$, there are points that lie within the circle $r=R$ but are not within $\Omega$.  This situation often arises when solving boundary value problems where the location of the boundary is perturbed~\cite{McCue2011,Farlow82}.  The key point being that the domain of $r$ in the functions $T(r,\theta)$ and $T_i(r,\theta)$ for $i=1,2,3,\ldots,$ is the same, $r < R(1+\varepsilon g(\theta))$.  However, the boundary value problems that define the terms in the perturbation solution, $T_i(r,\theta)$ for $i=1,2,3,\ldots,$ are defined on the original unperturbed domain, $r < R$.  Accordingly, our finite volume calculations and stochastic simulations are performed directly on the perturbed domain, $r < R(1+\varepsilon g(\theta))$,  whereas the perturbation solution is calculated on the unperturbed domain $r<R$ with the boundary conditions on $r=R(1+\varepsilon g(\theta))$ projected onto the unperturbed circle $r=R$. It is precisely this feature of the perturbation approach that allows us to construct such approximate solutions.  The same situation arises when we solve \eqref{eq:GovEq} on a perturbed ellipse, as we shall now explain. \cb

\subsection{Mean exit time on a perturbed ellipse} \label{perturbed_ellipse}
We proceed by deriving expressions for the exit time on a perturbed ellipse by considering $\partial \Omega$ as
\begin{align}\label{eq:PerturbEllipBnd1}
x& = a\left(1+\varepsilon g(\theta)\right)\cos \theta,\\ y &= b\left(1 + \varepsilon h(\theta)\right) \sin \theta,\label{eq:PerturbEllipBnd2}
\end{align}
where $2a > 0$ and $2b > 0$ are the width and height of the unperturbed ellipse, respectively, with $a > b$ and $g(\theta)$ and $h(\theta)$ are $\mathcal O(1)$ smooth periodic functions with period $2\pi$. Working in Cartesian coordinates, we suppose the solution takes the form
\begin{equation}\label{eq:PerturbEllipseAssump}
T(x, y) = T_0(x, y) + \epsilon T_1(x, y) + \epsilon^2 T_2(x, y) + \cdots + \epsilon^n T_n(x, y) + \mathcal O(\epsilon^{n+1}).
\end{equation}
To proceed, we impose Equation (\ref{eq:GovEqBoundary}) at the boundary specified in \eqref{eq:PerturbEllipBnd1} and (\ref{eq:PerturbEllipBnd2}) and expand about $\epsilon = 0$,
\begin{align}
0 &= T(a \cos \theta + a \epsilon g(\theta) \cos \theta, b \sin \theta + b \epsilon h(\theta) \sin \theta) \nonumber \\
0  & = T(a \cos \theta, b \sin \theta) + \sum_{k=1}^n \dfrac{1}{k!}\sum_{i=1}^k \binom{k}{i}\dfrac{\partial^kT}{\partial x^i\partial y^{k-i}} \left(a\epsilon g(\theta)\cos \theta\right)^i\left(b\epsilon h(\theta)\sin \theta\right)^{k-i} + \mathcal{O}(\epsilon^{n+1}),\label{eq:PerturbEllipseBndCnd}
\end{align}
where we evaluate the partial derivatives on the boundary of the unperturbed ellipse: $x = a \cos \theta$, and $y = b \sin \theta$.  From this point on in this Section we evaluate all partial derivatives like this on the boundary of the unperturbed ellipse.  For brevity it is convenient to define
\begin{align}
H_\ell(\theta) &= -\dfrac{\partial T_{\ell - 1}}{\partial y}bh(\theta)\sin \theta - \dfrac{\partial T_{\ell-1}}{\partial x}ag(\theta)\cos \theta  - \dfrac{1}{2}\dfrac{\partial^2 T_{\ell-2}}{\partial y^2}(bh(\theta)\sin \theta)^2  \nonumber \\
&- \dfrac{\partial^2 T_{\ell-2}}{\partial x \partial y}(a g(\theta)\cos \theta)(b h(\theta)\sin \theta) - \cdots - \dfrac{1}{\ell!}\dfrac{\partial^\ell T_0}{\partial x^\ell}(a g(\theta)\cos \theta)^\ell \nonumber \\\label{eq:Helltheta}
&= -\sum_{k=1}^\ell\sum_{i=0}^k\dfrac{1}{k!}\binom{k}{i}\frac{\partial^k T_{\ell-k}}{\partial x^i\partial y^{k-i}}(a g(\theta)\cos \theta)^i(bh(\theta)\sin \theta)^{k-i}.
\end{align}
Substituting \eqref{eq:PerturbEllipseAssump} into \eqref{eq:PerturbEllipseBndCnd}, and equating powers of $\epsilon$ gives
\begin{align}
\mathcal O(1) &: \quad T_0(a\cos \theta, b\sin \theta) = 0, \\
\mathcal O(\epsilon^\ell) &: \quad T_\ell(a \cos \theta, b\sin \theta) = H_\ell(\theta),
\end{align}
for $\ell=1,\ldots,n$. Substituting \eqref{eq:PerturbEllipseAssump} into \eqref{eq:GovEq} leads to the following family of boundary value problems
\begin{gather}\label{eq:EllipseBVPs1}
\mathcal O(1) : \quad D\nabla^2 T_0 = -1, \quad T_0\left(a \cos \theta, b\sin \theta \right) = 0,  \\
\mathcal O(\epsilon^\ell) : \quad \nabla^2 T_\ell = 0, \quad T_\ell(a \cos \theta, b \cos \theta) = H_\ell(\theta), \label{eq:EllipseBVPs2}
\end{gather}
for $\ell=1,\ldots,n$. The solution of \eqref{eq:EllipseBVPs1} is \eqref{eq:EllipseSolution},  and each higher order term is the solution of Laplace's equation on the ellipse with prescribed boundary data associated with the previous terms.  For example, the $\mathcal O(\epsilon^\ell)$ boundary value problem given in \eqref{eq:EllipseBVPs2} involves the boundary data $H_\ell(\theta)$, which depends on $T_{\ell-1}, T_{\ell-2},\ldots,T_{0}$ as evident in  \eqref{eq:Helltheta}. Following Jackson~\cite{Jackson1944} we construct the solution in terms of harmonic polynomials by expanding the boundary data $H_\ell(\theta)$ as a Fourier series,
\begin{gather}\label{eq:BoundaryDataFourier1}
H_\ell(\theta) = \dfrac{a_0}{2} + \sum_{k=1}^\infty \left(a_k \cos (k \theta) + b_k \sin(k\theta)\right), \quad 0 \leq \theta < 2\pi, \quad \textrm{where}, \quad \\
a_k = \dfrac{1}{\pi} \int_0^{2\pi} H_\ell(\theta) \cos(k\theta) \, \textrm{d}\theta, \quad b_k = \dfrac{1}{\pi} \int_0^{2\pi} H_\ell(\theta) \sin(k\theta) \,\textrm{d}\theta. \label{eq:BoundaryDataFourier2}
\end{gather}
We then compute $u_\ell(x, y) = \Re \left((x+iy)^\ell \right)$ and $v_\ell(x, y) = \Im\left((x+iy)^\ell \right)$ for $\ell=1,2,\ldots$, and also define $u_0(x, y) = 1$ and $v_0(x, y) = 0$, given explicitly  as
\begin{align}
u_\ell(x, y) &= \sum_{j=0}^{\left\lfloor \frac{\ell}{2}\right\rfloor}   \binom{\ell}{2j}x^{\ell-2j}(-1)^jy^{2j}, \\
v_\ell(x, y) &= \sum_{j=0}^{\left\lfloor \frac{\ell+1}{2}\right\rfloor} \binom{\ell}{2j+1}x^{\ell-2j-1}(-1)^jy^{2j+1}.
\end{align}
The next step is to compute $2\mathcal T_k(\cos\theta)$ for $k=1,2,\ldots$, where $\mathcal T_k$ is the Chebyshev polynomial of the first kind of degree $k$, and extract the coefficients of $\cos^r \theta$,  $C_r$, in this polynomial.  Using these expressions we compute, for even $k \geq 2$
\begin{align}
U_k(x, y) &= \dfrac{1}{(a+b)^k + (a-b)^k}\sum_{r=0}^{\frac{k}{2}} C_{2r}\left(a^2-b^2\right)^{\frac{k}{2}-r}u_{2r}(x, y), \\
V_k(x, y) &= \dfrac{1}{(a+b)^k - (a-b)^k}\sum_{r=0}^{\frac{k}{2}} C_{2r}\left(a^2-b^2\right)^{\frac{k}{2}-r} v_{2r}(x, y),
\end{align}
and for odd $k \geq 1$,
\begin{align}
U_k(x, y) &= \dfrac{1}{(a+b)^k + (a-b)^k}\sum_{r=0}^{\frac{k-1}{2}} C_{2r+1}\left(a^2-b^2\right)^{\frac{k-1}{2}-r}u_{2r+1}(x, y), \\
V_k(x, y) &= \dfrac{1}{(a+b)^k - (a-b)^k}\sum_{r=0}^{\frac{k-1}{2}} C_{2r+1}\left(a^2 - b^2\right)^{\frac{k-1}{2}-r}v_{2r+1}(x, y).
\end{align}
This gives us the solution of our $\mathcal O(\epsilon^\ell)$ problem,
\begin{equation}\label{eq:Ellipse_Laplace}
T_\ell(x, y) = \dfrac{a_0}{2} + \sum_{k=1}^\infty \left(a_kU_k(x, y) + b_kV_k(x, y)\right),
\end{equation}
where $a_0$, $a_k$ and $b_k$ are the Fourier coefficients from the boundary data, \eqref{eq:BoundaryDataFourier1}--Equation (\ref{eq:BoundaryDataFourier2}).

\begin{figure}[H]
	\centering
	\includegraphics[width=1.0\textwidth]{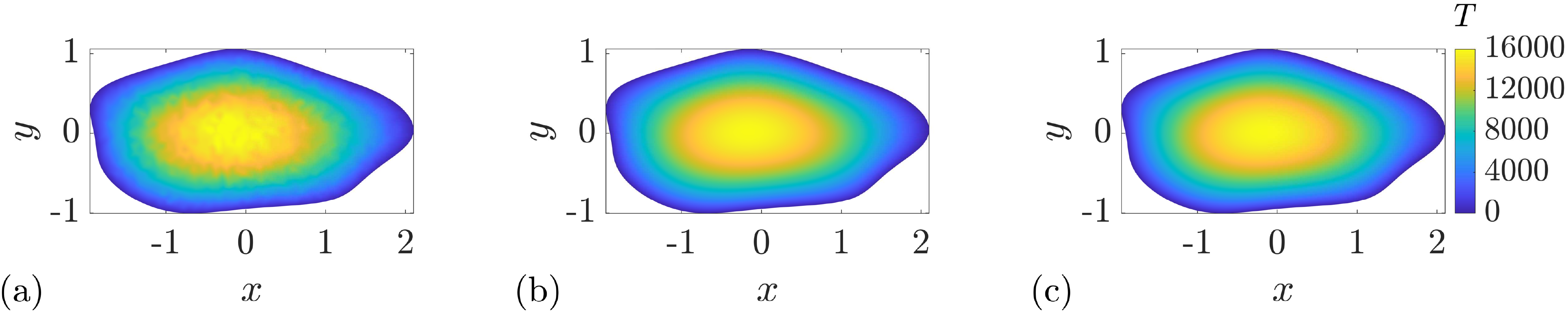}
	\caption{\textbf{Exit time on a perturbed ellipse.} (a) Averaged data from repeated stochastic simulations. (b) $\mathcal{O}(\varepsilon^2)$ perturbation solution. (c) Numerical solution of Equations (\ref{eq:GovEq})--(\ref{eq:GovEqBoundary}). Parameters are $\tau=1$, $\mathcal P=1$, $\delta = 1 \times 10^{-2}$ and $D=2.5 \times 10^{-5}$. The triangular mesh used to construct the solution in (a) and (c) has $1243$ nodes and $2342$ triangular elements. \cbl For (a), 1000 random walks starting from each node were generated. \cb}
	\label{fig:F4}
\end{figure}

This solution is straightforward to evaluate symbolically, and a \matlab\ implementation to evaluate it is available on \href{https://github.com/ProfMJSimpson/Exit_time}{GitHub}.  Results in Figure \ref{fig:F4}(a)--(c) provide a visual comparison of exit time data from repeated stochastic simulations, the perturbation solution and the numerical solution of \eqref{eq:GovEq} for a perturbed ellipse with $a=2$, $b=1$, $g(\theta) = \sin(3\theta) + \cos(5\theta) - \sin \theta$, $h(\theta) = \cos(3\theta) + \sin(5\theta) - \cos \theta$, and $\varepsilon = 1/20$.  The solutions in Figure \ref{fig:F4} show that the truncated perturbation solution can be very accurate, with just three terms in \eqref{eq:PerturbEllipseAssump}, and 25 terms in \eqref{eq:Ellipse_Laplace} required to produce a good match with the numerical solution.   Additional terms in the perturbation solution, or perturbation solutions for different choices of $\varepsilon$,  $g(\theta)$, or $h(\theta)$ can be evaluated using the \matlab\ algorithms on \href{https://github.com/ProfMJSimpson/Exit_time}{GitHub}.

\subsection{Case Studies} \label{sec:CaseStudies}
\begin{figure}[H]
	\centering
	\includegraphics[width=1.0\textwidth]{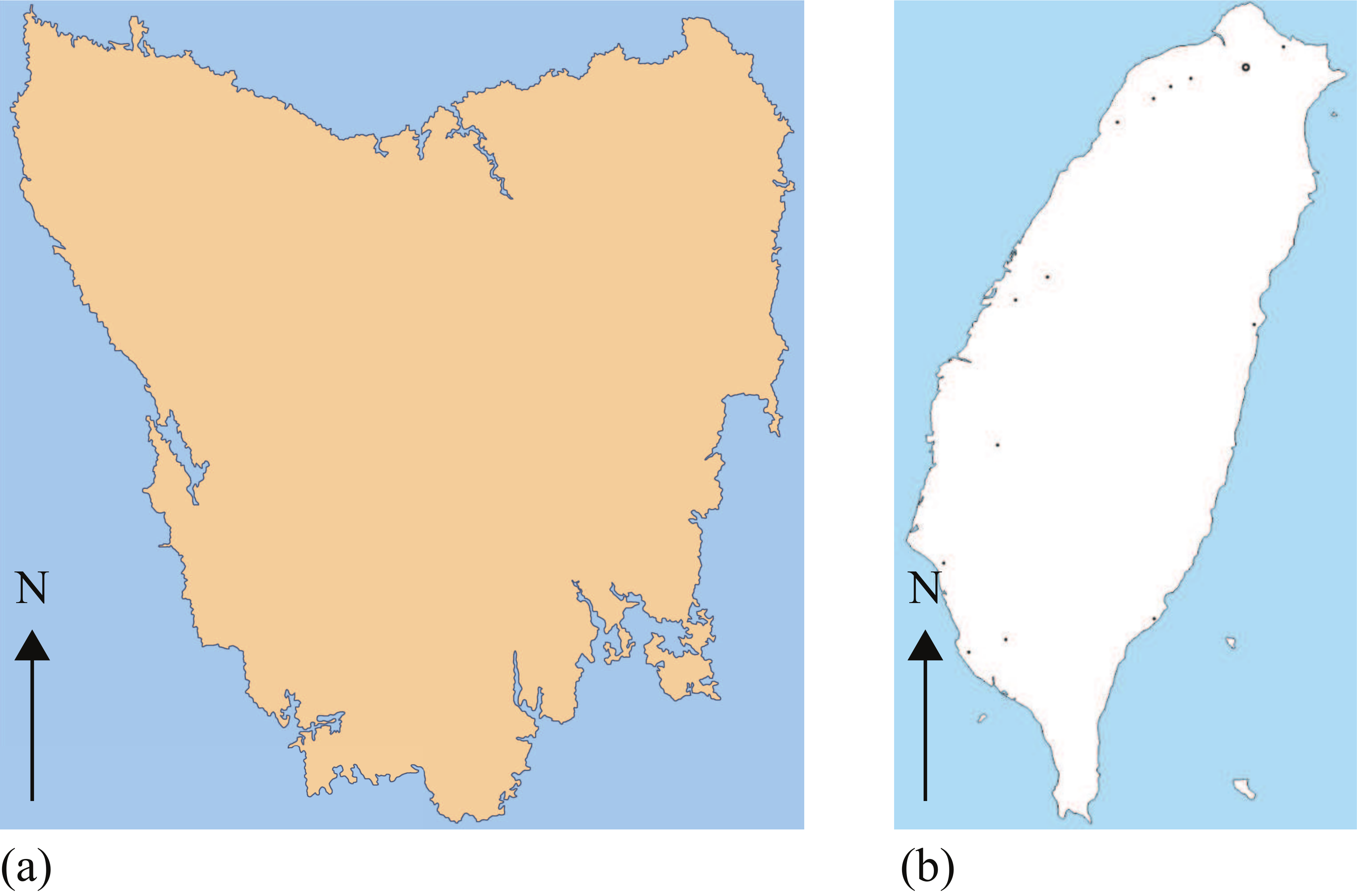}
	\caption{\textbf{Case studies.} (a) Tasmania~\cite{TassieMap}. (b) Taiwan~\cite{TaiwanMap}.}
	\label{fig:F5}
\end{figure}

To showcase how the perturbation solutions can be applied to naturally--occurring geometries, we now turn to the problem of taking an image of a region, approximating that region as a perturbed disc or perturbed ellipse, and then using our approach to estimate the exit time from that  region.  For this exercise we consider the islands of Tasmania and Taiwan.  In particular we represent the boundary of Tasmania as a perturbed disc and the boundary of Taiwan as a  perturbed ellipse, based on the maps in Figure \ref{fig:F5}.
Note that we have deliberately omitted any scale on the maps in Figure \ref{fig:F5} since we wish to focus on the role of shape rather than size. This allows us to represent the boundary of Tasmania as a perturbed unit disc, and to compare the exit time from this realistic geometry with the exit times from the more idealised shapes in Figure \ref{fig:F1} and Figure \ref{fig:F3}.   Similarly, we represent the boundary of Taiwan as a perturbed ellipse on a scale that allows us to compare the exit time distribution from this realistic geometry with the results in Figure \ref{fig:F2} and Figure \ref{fig:F4}. We now explain how we process the images in Figure \ref{fig:F5} to extract data that allows us to compute the exit times.

\begin{figure}[H]
	\centering
	\includegraphics[width=1.0\textwidth]{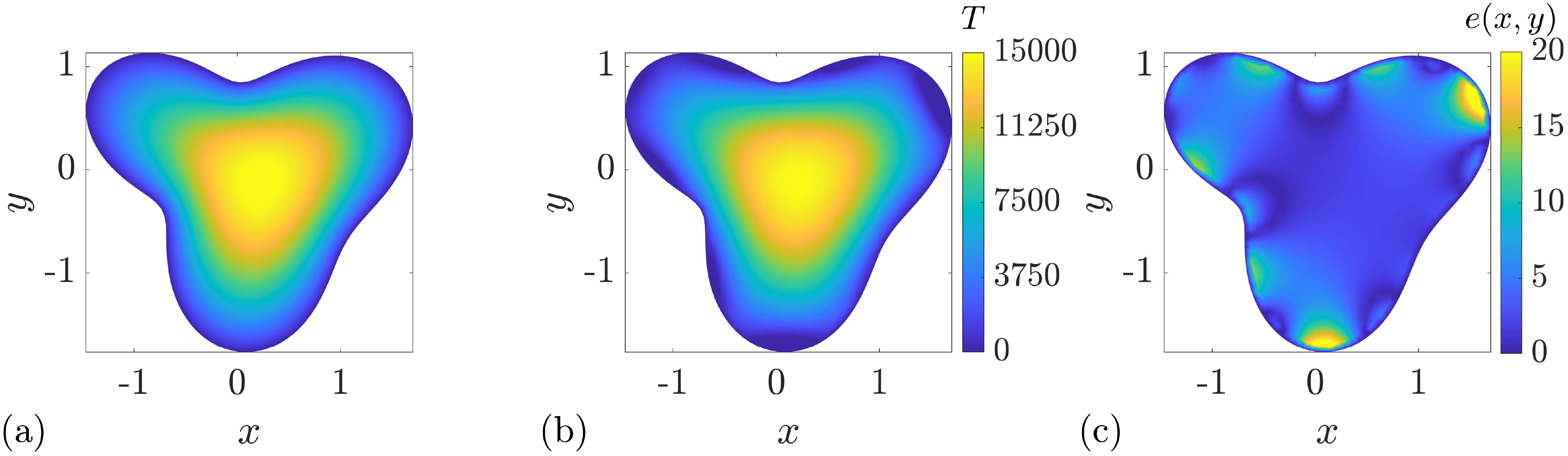}
	\caption{\textbf{Mean exit time on \cbl pseudo-Tasmania. \cb} (a) Numerical solution of Equations (\ref{eq:GovEq})--(\ref{eq:GovEqBoundary}). (b) $\mathcal{O}(\varepsilon^2)$ perturbation solution with $G=3$. (c) Discrepancy between the numerical and perturbation solution.  All results correspond to $D = 2.5 \times 10^{-5}$. The triangular  mesh used to construct the solution in (a) has $1188$ nodes and $2250$ triangular elements.}
	\label{fig:F6}
\end{figure}

To represent Tasmania as a perturbed disc we follow a two-step procedure.  First, we use \matlab\ image processing tools to describe the boundary of Tasmania as a set of points as described in the Appendix C, and we fit the disc $(x-x_c)^2 + (y-y_c)^2 = R^2$ to those points using a spline approximation in \matlab's \textit{cscvn}~\cite{MathworksCSCVN} function.  Second,  we shift this disc so that it is centered at the origin, and assume that the boundary takes the form $\mathcal{R}(\theta) = R(1 + \epsilon g(\theta))$, where we approximate $g(\theta)$ by
\begin{equation}
g(\theta) = A_0 + \sum_{n=1}^G \left(A_n \cos (n \theta) + B_n \sin(n\theta)\right).
\end{equation}
If the points $\{(x_i, y_i)\}_{i=1}^N$ represent the given boundary, we compute the polar angle for each point $\theta_i$. To represent our boundary in this way we require
\begin{equation}\label{eq:Tasmania_Boundary}
\frac{1}{R}\sqrt{x_i^2+y_i^2} - 1 = \epsilon A_0 + \epsilon\sum_{n=1}^G A_n \cos(n\theta_i) + \epsilon\sum_{n=1}^G B_n\sin(n\theta_i), \quad i=1,2,\ldots,N.
\end{equation}
We estimate the coefficients $A_{0}$, $A_{n}$, $B_{n}$ for $n = 1,2,\hdots,G$ by computing the least--squares solution of \eqref{eq:Tasmania_Boundary} that minimises the sum of squared residuals
\begin{equation}\label{eq:Tasmania_leastsquares}
\sum_{i=1}^{N}\left(\frac{1}{R}\sqrt{x_i^2+y_i^2} - 1 - \epsilon A_0 - \epsilon\sum_{n=1}^G A_n \cos(n\theta_i) - \epsilon\sum_{n=1}^G B_n\sin(n\theta_i)\right)^{2}.
\end{equation}
This least--squares solution is computed using \matlab's backslash operator. For simplicity we set $\epsilon = 1/10$ in this case study.  Given our estimates of the coefficients in \eqref{eq:Tasmania_Boundary}, our approximation of the boundary is shown in Figure \ref{fig:F6}, \cbl where we note that the approximation amounts to neglecting fine-scale features of the  Tasmanian coastline. For clarity we refer to this region as \textit{pseudo-Tasmania}. As we pointed out previously, our approach requires that $\mathcal{R}(\theta)$ is a single-valued function of $\theta$ such that any ray drawn from the origin intersects precisely one point of the boundary $\partial \Omega$.  This condition can only be met by neglecting the fine-scale structures of the boundary, especially at the South-East part of Tasmania where there is an obvious peninsula. \cb  Given this approximation, we apply the perturbation analysis in Section \ref{perturbed_disc} to give the results in Figure \ref{fig:F6}.
Figure \ref{fig:F6}(a) shows the numerical solution of \eqref{eq:GovEq} within the region enclosed by the boundary obtained by truncating  \eqref{eq:Tasmania_Boundary} with $G=3$ with boundary coordinate data obtained from the map in Figure \ref{fig:F5}(a).   All \matlab\ files required to replicate the boundary extraction and fitting are available on \href{https://github.com/ProfMJSimpson/Exit_time}{GitHub}.  Figure \ref{fig:F6}(b) shows the $\mathcal{O}(\varepsilon^2)$ perturbation solution, where infinite sums are approximated using the first 25 terms in \eqref{eq:DiscBVPSolutions}.  Visual comparison of the numerical and perturbation solutions indicates that the perturbation solution is remarkably accurate given that the domain in Figure \ref{fig:F6}(a)--(b) is quite far from a unit disc.  In fact, the maximum difference between the  boundary in Figure \ref{fig:F6}(a)--(b) and the underlying unit disc is, approximately, $0.64$, confirming that this domain is a reasonably large perturbation of a unit disc.   Careful comparison of the numerical and perturbation solutions show some discrepancy, particularly near the southern and north eastern portions of the boundary.

To quantify the discrepancy we introduce a measure of the difference between the two solutions,
\begin{equation}\label{eq:discrepancy}
e(x, y) = 100 \dfrac{\left|T_{\textrm{n}}(x, y) - T_{\textrm{p}}(x, y)\right|}{\max_{(x, y) \in \Omega} |T_{\textrm{n}}(x, y)|},
\end{equation}
where $T_{\textrm{n}}(x, y)$ is the numerical finite volume solution and $T_{\textrm{p}}(x, y)$ is the truncated perturbation solution, such that $e(x,y)$ is a convenient measure of the percentage relative error as a function of position.  The plot of $e(x,y)$ in Figure \ref{fig:F6}(c) confirms that the perturbation solution is remarkably accurate in the interior of the domain, with small discrepancies along some of the boundary.  While the small discrepancy along some parts of the boundary are visually discernable, these differences are not overwhelming, and the basic features of the numerical solution is captured by the perturbation solution.  More accurate perturbation solutions can be constructed by including more terms in the truncated perturbation solution, including more terms in the infinite sums, or both.  These options may be explored using the \matlab\ algorithms provided on \href{https://github.com/ProfMJSimpson/Exit_time}{GitHub}.  \cbl More details about the implications of approximating Tasmania by pseudo-Tasmania are given in Appendix D. \cb

\begin{figure}[H]
	\centering
	\includegraphics[width=1.0\textwidth]{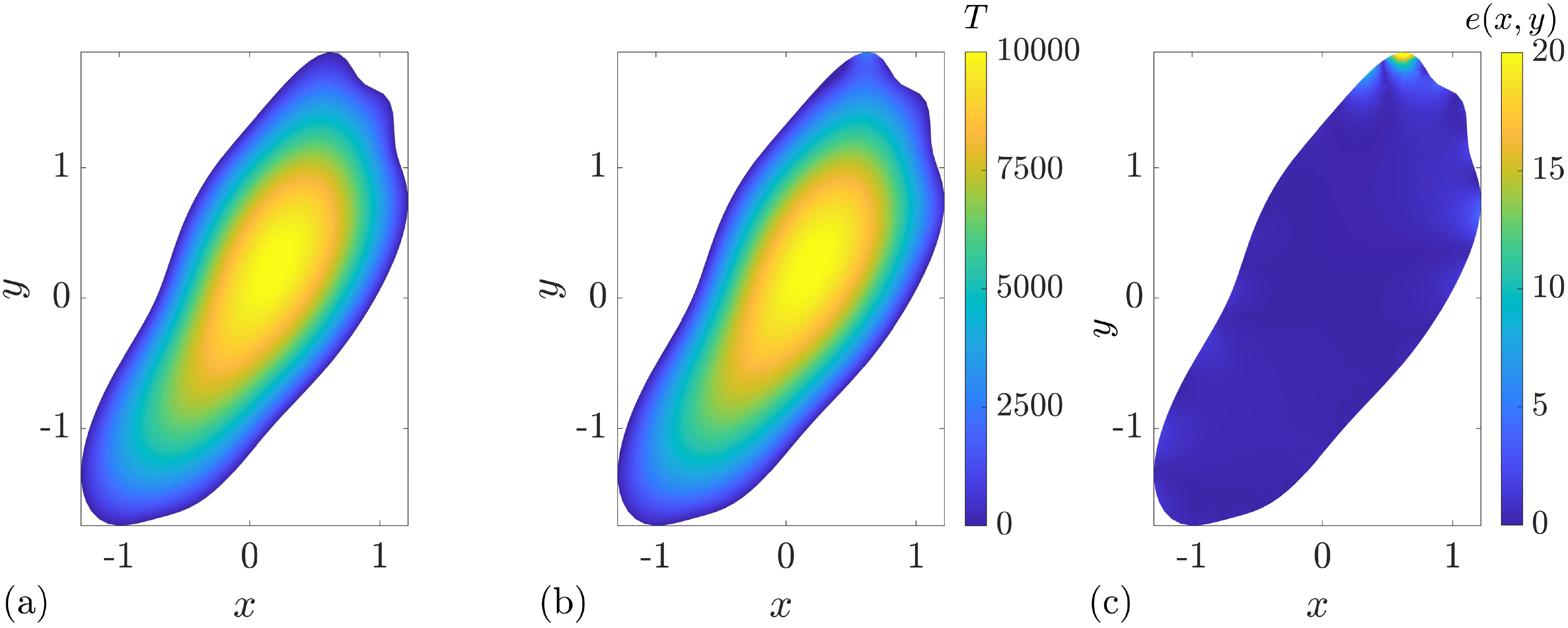}
	\caption{\textbf{Mean exit time on \cbl pseudo-Taiwan. \cb} (a) Numerical solution of Equations (\ref{eq:GovEq})--(\ref{eq:GovEqBoundary}). (b) $\mathcal O(\epsilon^2)$ perturbation solution. (c) Discrepancy between the numerical and perturbation solution. Parameters are $\tau = 1$, $\mathcal P=1$, $\delta = 1 \times 10^{-2}$, $D = 2.5 \times 10^{-5}$, $G = H =  9$. The triangular mesh used to construct the solution in (a) has $961$ nodes and $1803$ triangular elements.}
	\label{fig:F7}
\end{figure}

To represent Taiwan as a perturbed ellipse we first follow image pre-processing outlined in the Appendix C.  The method developed by Szpak et al.~\cite{Szpak2015} allows us to approximate the boundary as an ellipse with a particular orientation and centre.  We then rotate and shift this identified ellipse so that it is centered at the origin with semi--major axis along the $x$--axis.  To approximate the boundary of Taiwan as a perturbed ellipse,  \eqref{eq:PerturbEllipBnd1} and (\ref{eq:PerturbEllipBnd2}), we represent the functions $g(\theta)$ and $h(\theta)$ as
\begin{align} \label{eq:Taiwan_Boundary1}
g(\theta) & =  A_0 + \sum_{n=1}^G \left(A_n \cos (n\theta) + B_n\sin(n\theta)\right), \\
h(\theta) &= C_0 + \sum_{n=1}^H \left(C_n \cos(n\theta) + D_n\sin(n \theta)\right). \label{eq:Taiwan_Boundary2}
\end{align}
As before, the boundary is given by a set of points, $\{(x_i, y_i)\}_{i=1}^N$, and we compute the polar angle $\theta_i$ for each point. Representing the boundary using this approach leads to two systems of linear equations
\begin{align}
\frac{x_i}{a\cos\theta_i} - 1 &= \epsilon A_0 + \epsilon \sum_{n=1}^G A_n\cos(n\theta_i) + \epsilon \sum_{n=1}^H B_n\sin(n\theta_i),\quad i=1,2,\ldots,N,\label{eq:overdet1}\\
\frac{y_i}{b\sin\theta_i} - 1 &= \epsilon C_0 + \epsilon \sum_{n=1}^H  C_n \cos(n\theta_i) + \epsilon \sum_{n=1}^H D_n\sin(n\theta_i),\quad i=1,2,\ldots,N\label{eq:overdet2}.
\end{align}
As for Tasmania, we estimate the coefficients, $A_{0}$, $A_{n}$, $B_{n}$ for $n=1,2\ldots,G$ in \eqref{eq:overdet1} and $C_{0}$, $C_{n}$, $D_{n}$ for $n=1,2\ldots,H$ in \eqref{eq:overdet2} by computing the least--squares solution of each linear system using MATLAB's backslash operator. In summary, we can represent the boundary of Taiwan using $g(\theta)$ and $h(\theta)$ given by \eqref{eq:PerturbEllipBnd1} and \ref{eq:PerturbEllipBnd2} for some choice of $\epsilon$, which we again take to be $\epsilon = 1/10$ in this case study.  Figure \ref{fig:F7}(a) shows the numerical solution of \eqref{eq:GovEq} within the region enclosed by the boundary obtained by truncating  \eqref{eq:Taiwan_Boundary1}--(\ref{eq:Taiwan_Boundary2}) with $G=H=9$ that is based on boundary data obtained from the map in Figure \ref{fig:F5}(b).  All \matlab\ files required to replicate the boundary extraction and fitting are available on \href{https://github.com/ProfMJSimpson/Exit_time}{GitHub}.  The solution in Figure \ref{fig:F7}(b) is the $\mathcal{O}(\varepsilon^2)$ perturbation solution.  Visual comparison of the numerical and perturbation solutions indicates that the perturbation solution is remarkably accurate, and the plot of $e(x,y)$, given by \eqref{eq:discrepancy} in Figure \ref{fig:F7}(c) confirms this accuracy, even along the boundaries. \cbl Again, this accuracy is obtained through neglecting the very fine-scale features of the coastline of Taiwan that would never be accurately represented by a perturbed ellipse. \cb

\section{Conclusions and outlook} \label{Conc}
In this work we consider the canonical problem of determining the mean first passage time for diffusion, which requires the solution of an elliptic partial differential equation on the domain of interest.  This problem has been studied, in detail, both analytically and computationally, with many known exact solutions for relatively simple geometries, such as lines, discs and spheres~\cite{Redner2001,Krapivsky2010,Hughes1995}.  The calculation of exact expressions for the mean first passage time for more complicated geometries is an active, and ongoing field of research.  In this work we present new solutions for the mean first passage time for diffusion on irregular two--dimensional domains, where these solutions are obtained in terms of a perturbation of the classical results for the mean first passage time on a disc or an ellipse.  The expressions we derive for perturbed discs and perturbed ellipses are tested using numerical solutions of the governing partial differential equation. We show that the perturbation solutions rapidly converge to the numerical solution with just a small number of terms that are straightforward to evaluate using \matlab\ code supplied on \href{https://github.com/ProfMJSimpson/Exit_time}{GitHub}. Finally, we show how to estimate the exit time in naturally--occurring shapes by representing the boundaries of Tasmania and Taiwan as a perturbed disc and ellipse, respectively, and then evaluating the exit time on these shapes.

There are many avenues available to extend the work presented here.  Here we consider the most fundamental transport mechanism: unbiased diffusion, but it is also possible to consider generalisations of \eqref{eq:GovEq} such as drift--diffusion,  diffusion--decay~\cite{Ellery2012a,Ellery2012b} or more complicated discrete mechanisms including L\'evy flights~\cite{Wardak2020,Padash2020}.  A further extension is to consider calculating both the mean first passage time and higher moments of exit time~\cite{Carr2018}.  All problems in this work consider exit times by specifying absorbing boundary conditions in the random walk model, which correspond to homogeneous Dirichlet conditions in the partial differential equation model.  These can be extended to mixed boundary conditions where some parts of $\partial \Omega$ are absorbing, and other parts of $\partial \Omega$ are reflecting on the original, unperturbed boundary.  It would then be very interesting to consider perturbations with such mixed boundary conditions.  \cbl A more substantial extension of this work would be to consider the solution of \eqref{eq:GovEq} on more complicated shapes that are not small, smooth perturbations of a disc or an ellipse. If, for example, we consider the case where $\Omega$ corresponds to a larger circle whose boundary just touches a smaller circle, we are unable to directly apply the techniques developed in this work, and so a different approach is required to construct approximate solutions of \eqref{eq:GovEq}. \cb  In addition to the more mathematical extensions described here, it is also possible to consider extensions of the present work that are more computational.  For example, further consideration could be given to the way in which the perturbation solutions presented in this work are evaluated.  In this work we evaluate the perturbation solutions using symbolic tools in \matlab\ since it is convenient for us to provide a single software to perform stochastic random walk simulations, finite volume numerical calculations and to evaluate the perturbation solution in a single programming language.  We note, however, that working with a different symbolic language could be more efficient, especially if additional terms in the perturbation solution are to be evaluated.

\noindent
\textit{Acknowledgements:} This work is supported by the Australian Research Council (DP200100177) and Queensland University of Technology for providing a summer research scholarship to DJV.  \cbl We thank two referees for helpful suggestions. \cb

\newpage
\section*{Appendix A: Numerical methods}
\subsection*{Stochastic simulations}
We simulate particle lifetime distributions using continuous space, discrete time random walk stochastic simulations.  Time is discretized with constant time steps of duration $\tau > 0$.  In each time step, a particle, at location $\mathbf{x}(t)$, attempts to step a distance $\delta > 0$, to $\mathbf{x}(t + \tau)= \mathbf{x}(t) + \delta(\cos\theta,\sin\theta)$ with probability $\mathcal{P} \in [0,1]$.  Here, $\theta$ is sampled from a uniform distribution, $\theta \sim \mathcal{U}[0, 2\pi]$.  This discrete process corresponds to a random walk with diffusivity $D = \mathcal{P} \delta^2/(4 \tau)$~\cite{Hughes1995}.   Simulations continue until the particle steps across the boundary of the domain, and the exit time is recorded.  All simulations in this work use $\tau=1$, $\mathcal P=1$ and $\delta = 1 \times 10^{-2}$, giving $D=2.5 \times 10^{-5}$.

To estimate the mean exit time we consider $N$ identically--prepared realisations of the discrete process with starting position $\mathbf{x}(0)=(x,y)$.  This gives us $N$ estimates of the exit time from which we calculate the mean, $\displaystyle{T_{\textrm{sim}}(x,y) =(1/N) \sum_{i=1}^{N} t_{i}}$, where $t_i$ is the exit time from the $i$th identically prepared stochastic realisation.  In practice we use the stochastic model to estimate $T_{\textrm{sim}}(x,y)$ with $N=1 \times 10^3$ simulations, and these estimates are obtained at a number of spatial points in $\Omega$.  We consider an unstructured triangular meshing of $\Omega$, and we estimate $T_{\textrm{sim}}(x,y)$ at each node.  This means that for a meshing with $M$ nodes, we perform a total of $M \times N$ stochastic simulations.  For example, the results in Figure \ref{fig:F1}(a) are generated with $N=1000$ identically--prepared realisations at each of the $M=632$ nodes, giving a total of $632000$ stochastic simulations.  The resulting point estimates of $T_{\textrm{sim}}(x,y)$ at each node are interpolated across $\Omega$ using the \textit{interp} option in MATLAB's \textit{shading} function~\cite{MathworksSHADING} to provide a continuous estimate of the distribution of the mean exit time.

\begin{figure}[H]
	\centering
	\includegraphics[width=1.0\textwidth]{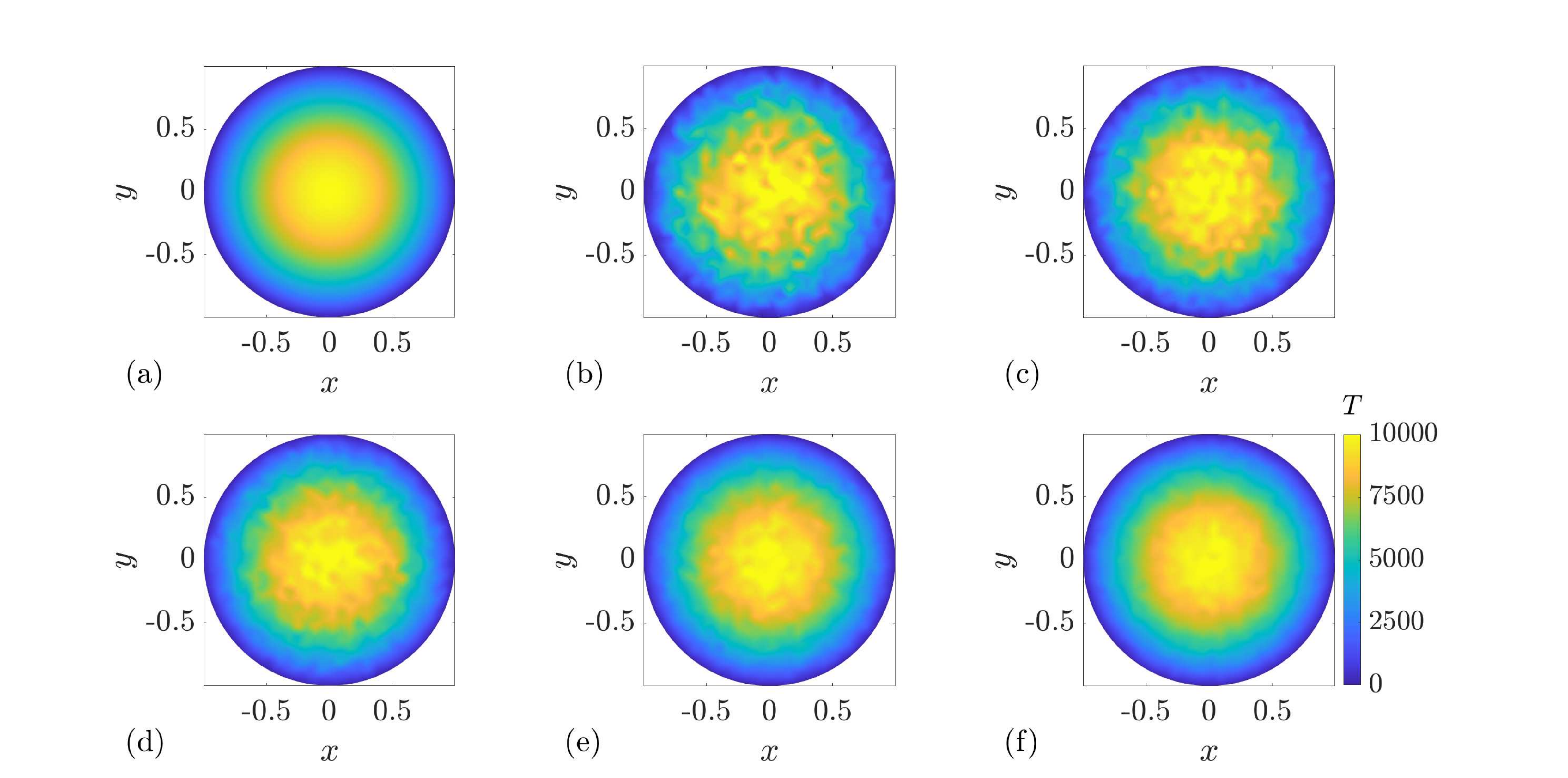}
	\caption{\cbl \textbf{Stochastic simulations as $N$ increases.} (a) Exact solution for the mean exit time on the unit disc, \eqref{eq:Solutiondisc}. (b)--(f) averaged simulation data for the mean exit time generated using $N=60, 125, 250, 500$ and $1000$ particles released per node in the finite volume mesh. \cb}
	\label{fig:FS1}
\end{figure}

\begin{figure}[H]
	\centering
	\includegraphics[width=1.0\textwidth]{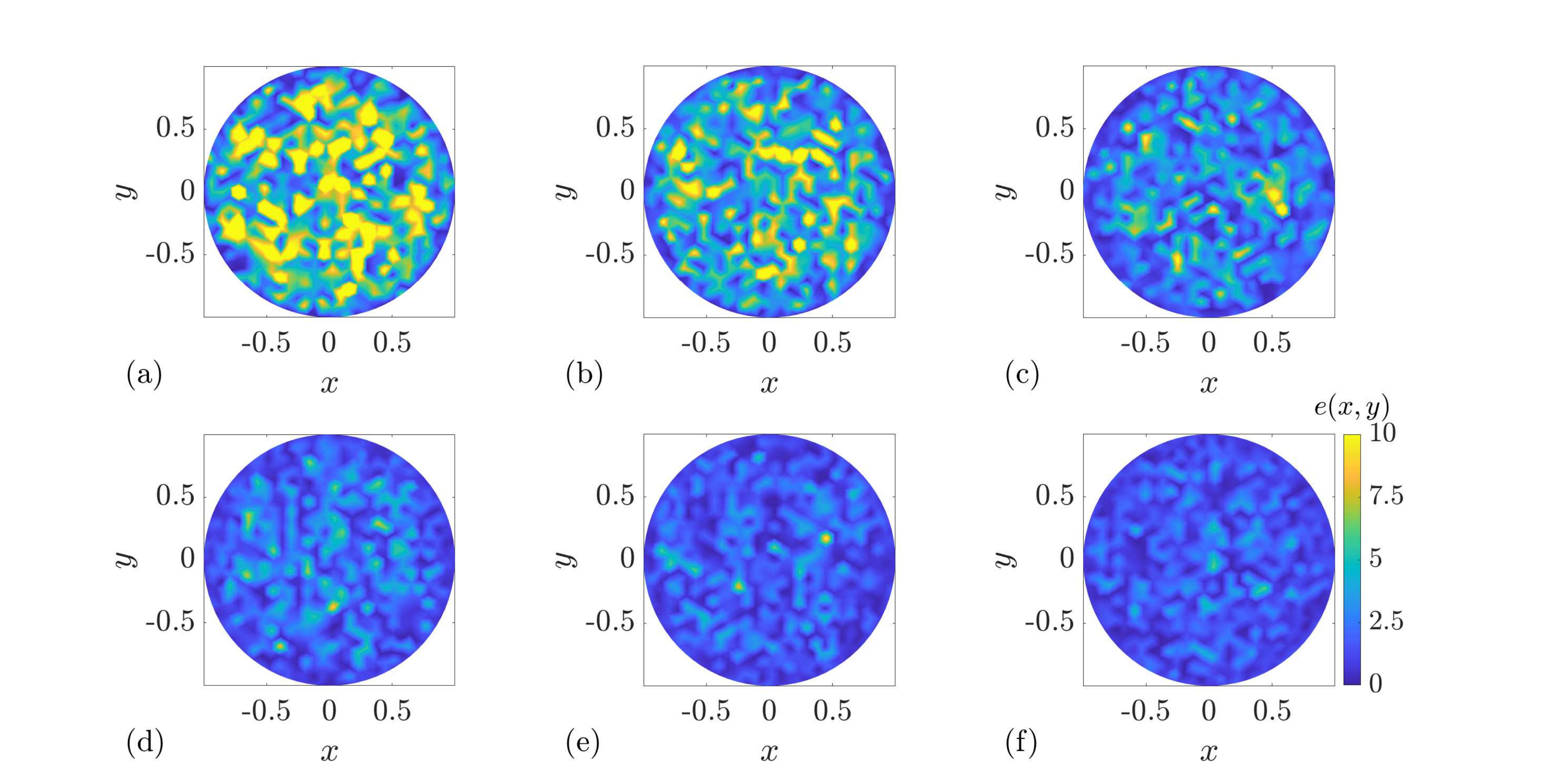}
	\caption{\cbl  \textbf{Comparison of stochastic simulations and exact solution as $N$ increases.} (a)--(f) Comparison of the exact solution for the mean exit time on the unit dist with simulation data for $N=60, 125, 250, 500, 750$ and $1000$, respectively.  All results are presented in terms of $e(x,y)$, given by \eqref{eq:discrepancy}. \cb}
	\label{fig:FS2}
\end{figure}

\cbl Results in Figure \ref{fig:FS1} provide a visual comparison of the impact of varying $N$.  The solution in Figure \ref{fig:FS1}(a) is the exact solution of \eqref{eq:GovEq} on the unit disc, \eqref{eq:Solutiondisc}. Data in Figure \ref{fig:FS1}(b)--(f) show estimates of the mean exit time on the unit disc generated using $N=60, 125, 250, 500$ and $1000$ particles per node in the finite volume mesh.  Comparing data in Figure \ref{fig:FS1}(b)--(f) we see that the fluctuations in the estimates of $T_{\textrm{sim}}(x,y)$ appear to decrease, as expected, as $N$ increases.  Additional results in Figure \ref{fig:FS2} compares the exact solution and simulation-based estimates as a function of $N$ in terms of \eqref{eq:discrepancy}. Again, we see that $e(x,y)$ approaches zero as $N$ increases.  These results justify our choice of $N=1000$ in the main document since we roughly have $e(x,y) < 5\%$ for $N=1000$.  Of course the algorithms available on \href{https://github.com/ProfMJSimpson/Exit_time}{GitHub} can be used to generate $T_{\textrm{sim}}(x,y)$ for larger $N$, but this comes with the drawback that increasing $N$ leads to longer simulation times.   \cb

\subsection*{Finite volume calculations}
We solve \eqref{eq:GovEq} numerically using a finite volume approximation~\cite{Eymard2000} to discretize the governing equation over an unstructured triangular meshing of $\Omega$.  To perform these calculations we use mesh generation software, GMSH~\cite{Geuzaine09}. The finite volume method is implemented using a vertex centered strategy with nodes located at the vertices in the mesh.  Control volumes are constructed around each node by connecting the centroid of each triangular element to the midpoint of its edges~\cite{Carr2016}. Linear finite element shape functions are used to approximate gradients in each element. Assembling the finite volume equations yields a sparse linear system that can be stored and solved efficiently.  For each numerical solution reported in this work we report the number of nodes and elements in the finite volume mesh, and in each case use a prescribed mesh element size of $0.08$ in GMSH~\cite{Geuzaine09} to generate these meshes. A \matlab\ implementation of the numerical algorithm is available on \href{https://github.com/ProfMJSimpson/Exit_time}{GitHub}.

\newpage
\subsection*{Appendix B: Truncation effects}
Results in Figures \ref{fig:F3}--\ref{fig:F4} compare estimates of mean exit time, $T$, for a perturbed disc and sphere, respectively.  In these figures we compare $T$ from repeated stochastic simulations, a truncated perturbation solution and a fine-mesh finite volume solution of the governing boundary value problem.  In these comparisons we choose a particular truncation of the perturbation solution to ensure that the perturbation solution and the finite volume solutions compare reasonably well.  Here, we explore how the choice of truncation impacts the accuracy of the perturbation solutions. The comparison between the perturbation and finite volume solutions in Figures \ref{fig:F3}--\ref{fig:F4} correspond to $\mathcal{O}(\varepsilon^2)$ perturbation solutions with 25 terms retained in the truncated infinite sum.  The visual comparison between the perturbation and finite volume solutions in Figures \ref{fig:F3}--\ref{fig:F4} indicates that the perturbation solution is very accurate.

\begin{figure}[H]
	\centering
	\includegraphics[width=1.0\textwidth]{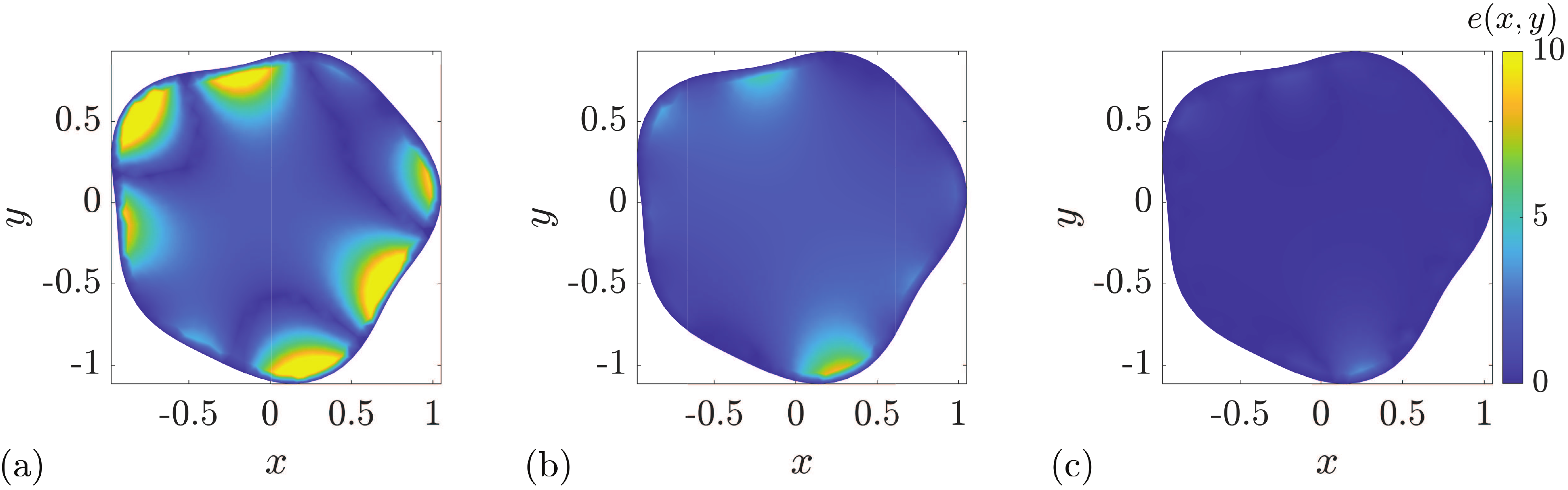}
	\caption{\textbf{Perturbation solution on a perturbed disc.} Comparison of perturbation solution to the finite volume approximation of the exit time on the perturbed disc, using: (a) $\mathcal{O}(\varepsilon)$ perturbation solution with 1 term in the truncated infinite sum; (b) $\mathcal{O}(\varepsilon)$ perturbation solution with 10 terms in the truncated infinite sum; and, (c) $\mathcal{O}(\varepsilon^2)$ perturbation solution with 10 terms in the truncated infinite sum.  All results are presented in terms of $e(x,y)$, given by \eqref{eq:discrepancy}.}
	\label{fig:FS3}
\end{figure}

\begin{figure}[H]
	\centering
	\includegraphics[width=1.0\textwidth]{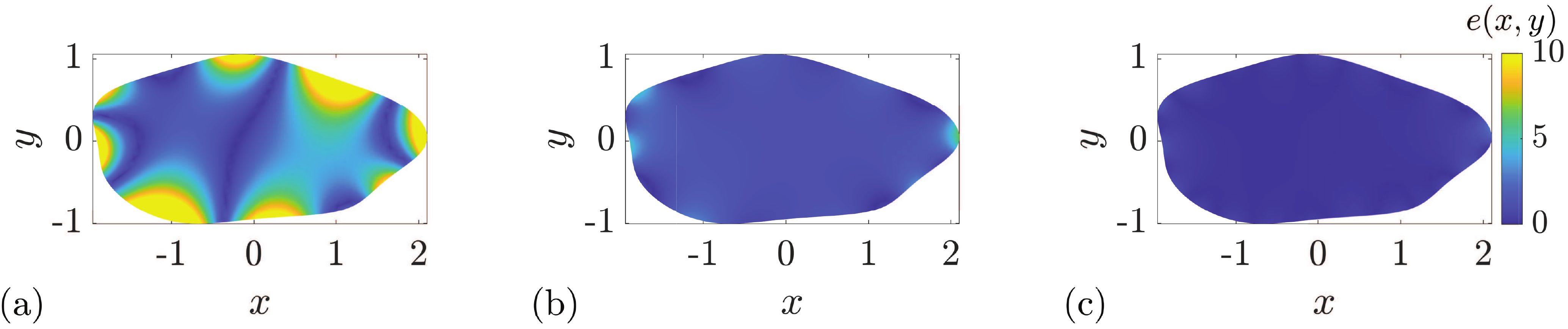}
	\caption{\textbf{Perturbation solution on a perturbed ellipse.} Comparison of perturbation solution to the finite volume approximation of the exit time on the perturbed ellipse, using: (a) $\mathcal{O}(\varepsilon)$ perturbation solution with 1 term in the truncated infinite sum; (b) $\mathcal{O}(\varepsilon)$ perturbation solution with 10 terms in the truncated infinite sum; and, (c) $\mathcal{O}(\varepsilon^2)$ perturbation solution with 10 terms in the truncated infinite sum.  All results are presented in terms of $e(x,y)$, given by \eqref{eq:discrepancy}.}
	\label{fig:FS4}
\end{figure}

We now quantify this comparison using \eqref{eq:discrepancy}.  Additional results in Figure \ref{fig:FS3} show plots of $e(x,y)$ for the same problem examined in Figure \ref{fig:F3}, except that here we use different truncations in the perturbation solution.   Comparing results in Figure \ref{fig:FS3}(a)--(c) indicate that the perturbation solution rapidly approaches the finite volume solution using only a modest number of terms.  Note that the perturbation solution in Figure \ref{fig:F3} is even more accurate then those in Figure \ref{fig:FS1}.  A similar comparison in Figure \ref{fig:FS4} confirms that the perturbation solution for the ellipse also rapidly approaches the numerical solution as with only a small number of terms.  Results in Figure \ref{fig:FS4} correspond to the problem previously explored in Figure \ref{fig:F4}.

\newpage
\subsection*{Appendix C: Image processing}
To apply our analysis to the boundary of Tasmania we use \matlab\ to produce an array representation of the boundary, and we smooth some of the boundary by refining certain jagged edges and the removal of some small peninsulas. After smoothing, we use the Sobel edge detection method in \matlab's \textit{cscvn}~\cite{MathworksCSCVN} function and the \textit{imclose}~\cite{MathworksIMCLOSE} function to detect and refine the boundaries. Boundary points on the detected edges are obtained with \textit{bwboundaries}~\cite{MathworksBWBOUNDARIES}.  Given a relatively dense set of points along the boundary, we retain each $30$th point to give the boundary in Figure \ref{fig:F6}. We compute the mean of the $x$ and $y$ coordinates and shift the data so that the centre of the region is at the origin.  We then scale the data so that it is comparable to a unit disc~\cite{Szpak2015}.  After numerical and perturbation solutions are obtained on the region contained in this boundary we rotate the resulting solutions to match the shape of the original region.

To apply our analysis to the boundary of Taiwan we start by manually smoothing some jagged portions of the boundary, and then use the Sobel method with \textit{imclose} and \textit{bwboundaries}~\cite{MathworksIMCLOSE, MathworksBWBOUNDARIES} to represent the boundaries in terms of a dense set of points.   We work with every $40$th point, and shift the region so that the centroid is at the origin, followed by a counterclockwise rotation of $\pi/3$ so that the semi--major axis co-insides with the $x$--axis, allowing us to apply the results in Section \ref{perturbed_ellipse}. The data is then scaled so that it is comparable to an ellipse with semi-major axis $2$ and semi-minor axis $1$~\cite{Szpak2015}.  We now apply the procedure described in Section \ref{sec:CaseStudies}~\cite{Szpak2015} to identify the best--fitting ellipse, and we then use the least--squares procedure described in Section \ref{sec:CaseStudies} to calculate  trigonometric polynomial representations of $g(\theta)$ and $h(\theta)$.

\newpage
\cbl
\subsection*{Appendix D: Comparing Tasmania and pseudo-Tasmania}
\begin{figure}[H]
	\centering
	\includegraphics[width=1.0\textwidth]{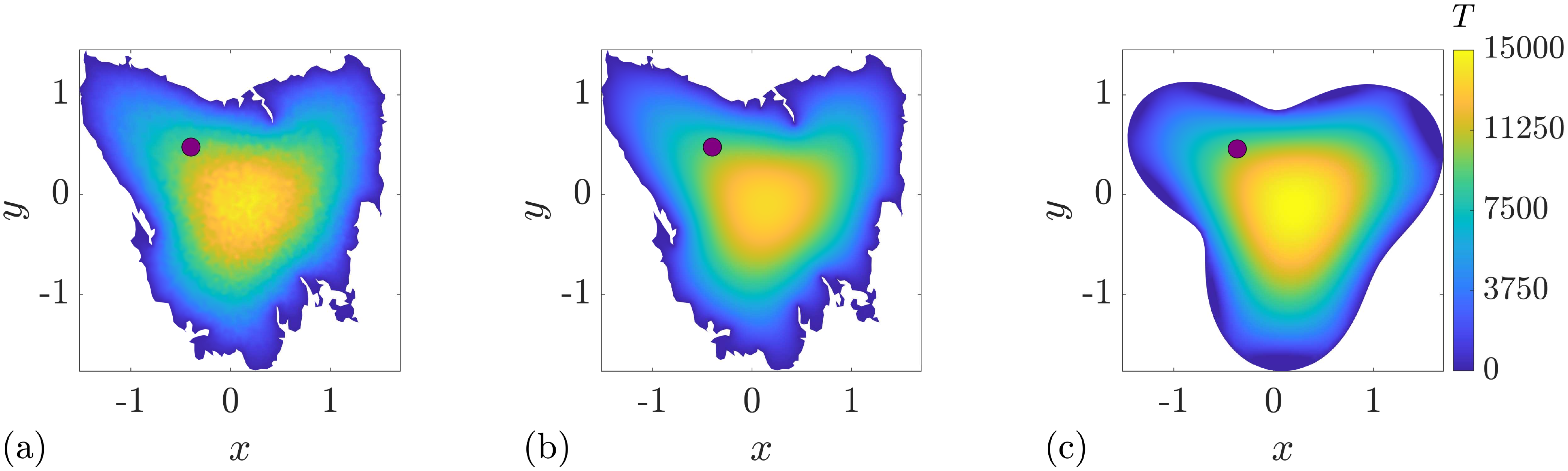}
	\caption{\textbf{Comparing Tasmania and pseudo-Tasmania.} (a) Averaged data from repeated stochastic simulations on Tasmania. (b) Numerical solution of Equations (\ref{eq:GovEq})--(\ref{eq:GovEqBoundary}) on Tasmania. (c) Exit time on pseudo-Tasmania using the perturbation solution from Figure 6(b). On each map we show the approximate location of Cradle mountain (purple disc). The triangular mesh used to construct the solution in (a) and (b) has $3356$ nodes and $6369$ triangular elements. For (a), 1000 random walks starting from each note were generated.}
	\label{fig:FS5}
\end{figure}

To quantitatively examine the implications of smoothing the boundaries of the natural coastlines we show additional results in Figure \ref{fig:FS5} where we compare exit times on Tasmania and pseudo-Tasmania.  Results in Figure \ref{fig:FS5}(a)-(b) show the exit time on Tasmania using repeated stochastic simulations and the finite volume numerical solution of Equations (\ref{eq:GovEq})--(\ref{eq:GovEqBoundary}), respectively. Figure \ref{fig:FS5}(c) repeats the perturbation solution on pseudo-Tasmania shown previously in Figure \ref{fig:F6}(b).  Visual inspection of the solutions on Tasmania and pseudo-Tasmania in Figure \ref{fig:FS5} indicate that the solutions compare very well in the interior of the domain, with the key difference being along the coastlines, as one might anticipate.   To provide a more quantitative comparison we consider the inland location of Cradle Mountain, shown in Figure \ref{fig:FS5} as a purple disc.  Our results give $T_{\textrm{sim}}=9427$ using repeated stochastic simulations on Tasmania, whereas we obtain $T_{\textrm{p}}=9673$ on pseudo-Tasmania, giving a discrepancy of just $2.6\%$.

\cb

\end{document}